# Properties of CdTe layers deposited by a novel method -Pulsed Plasma Deposition

C. Ancora[1], P. Nozar[3,*], G. Mittica[2], F. Prescimone[1], A. Neri[2], S. Contaldi[4], S. Milita[4], C. Albonetti[3], F. Corticelli[4], A. Brillante[5], I. Bilotti[5], G. Tedeschi[1] and C. Taliani[2,3]

[1] Siena Solar Nanotech, S.p.A., Piazza dell'Abbadia, 4, 53100 Siena, Italy,
[2] Organic Spintronics, S.r.l., via Gobetti 52/2, 40129 Bologna, Italy
[3] Istituto per lo Studio dei Materiali Nanostrutturati, via Gobetti, 101, 40129 Bologna, Italy,
[4] Istituto per la Microelettronica e i Microsistemi, via Gobetti, 101, 40129 Bologna, Italy
[5] Dipartimento di Chimica Fisica e Inorganica, Viale del Risorgimento, 4, 40136 Bologna, Italy

(*) Corresponding author: Tel.: +390516398503 Fax: +390516398540, E-Mail address: p.nozar@bo.ismn.cnr.it

## Abstract

CdTe and CdS are emerging as the most promising materials for thin film photovoltaics in the quest of the achievement of grid parity. The major challenge for the advancement of grid parity is the achievement of high quality at the same time as low fabrication cost. The present paper reports the results of the new deposition technique, Pulsed Plasma Deposition (PPD), for the growth of the CdTe layers on CdS/ZnO/quartz and quartz substrates. The PPD method allows to deposit at low temperature. The optical band gap of deposited layers is 1.50 eV, in perfect accord with the value reported in the literature for the crystalline cubic phase of the CdTe.

The films are highly crystalline with a predominant cubic phase, a random orientation of the grains of the film and have an extremely low surface roughness of 4.6±0.7 nm r.m.s.. The low roughness, compared to traditional thermal deposition methods (close space sublimation and vapour transport) permits the reduction of the active absorber and n-type semiconductor layers resulting in a dramatic reduction of material usage and the relative deposition issues like safety, deposition rate and ultimately cost

## 1. Introduction

Presently the dominant materials for thick solar cells realization are single crystalline and polycrystalline silicon. The disadvantage of these materials is their indirect optical band gap, which requires large thickness of the high purity active layer (hundreds of micrometers) to assure sufficiently high efficiency of the device. The large thickness of the silicon layer and the large energy required to produce the desired degree of the material purity results into high consumption of the prime material and to unsustainable high final cost of the resulting solar cell.

Grid parity imperative requires the application of thin film solar cells which adopt active material layers exhibiting direct optical band gap like CdTe and CIGS (Copper Indium Gallium Disulfide or Diselenide). As the consequence of the direct optical band gap these materials require substantially thinner active layer for the total absorption of the incident light. Thin film technology results in significant less consumption of the starting materials and cheaper production technology with respect to the silicon solar cells.

CdTe is becoming the most successful contender to silicon photoactive material for the realization of solar cells due to its relative simplicity of the composition and the structural and compositional stability (with respect, e.g. CIGS). However, the preparation of this material requires high deposition temperature to achieve the crystalline CdTe and the CdTe/CdS mixture at the interface. Moreover the poor surface quality prevents the use of films thinner than 6 μm. Both these drawbacks affect adversely the final cost. There is therefore a strong request for new efficient deposition techniques. The PPD technique is able to satisfy all this conditions and, therefore, is a method of choice for the industrial production of wide area thin films solar cells.

There are several different preparation methods used for the deposition of CdTe. Close Space Sublimation (CSS) and Vapour Transport Deposition (VTD) are already industrially exploited deposition techniques. Both methods suffer the common disadvantages, e.g. high deposition temperature (thus, the dependence on substrate properties) [1], high film surface roughness [2], high starting material consumption. Magnetron-Sputtering, and Pulsed Laser Deposition (PLD) are deposition techniques giving much better results, but, due to prohibitively high running costs, low production rate, strongly limited deposition area and difficulty to achieve the desired stoichiometry [3,4], are suitable for laboratory applications only. All these techniques have demonstrated some advantages and some important drawbacks with respect to the quality of the deposited CdTe thin films.

A new emerging and very promising thin film deposition technology is now revolutionizing the field of thin film preparation methods. This method is the Pulsed Plasma Deposition (PPD) which is being developed by Organic Spintronics, Srl. [5-7].

The PPD method properly adapted and tuned for photoactive materials is the technology based on the production of the pulsed plasma accompanied by the generation of high energy electron pulses (up to 20 keV) directed to the source of deposited material (target). Electron and plasma beam interaction with target generates the target material plasma which propagates towards the substrate and deposits on its surface.

This technique, due to simple and compact electron gun construction and high deposition rates, can be easily adapted to large area deposition and represents a new choice for industrial production of thin films for not only photovoltaic applications.

## 2. Experimental

The core of the PPD system is an electron gun generating the pulsed beam of plasma and high-energy electrons (up to 20 keV). The plasma and electron beam is created from working gas (e.g. oxygen, argon or nitrogen) flowing at low pressures ($10^{-6}$ - $10^{-2}$ mbar) through the gun operational body. The electron gun body is mainly composed by metallic hollow cathode and relatively narrow (2-4 mm diameter) dielectric channel (typically 80-150 mm long) or capillary.

Electrons and plasma generated inside the hollow cathode are collected and accelerated by an electrical potential difference (up to 20 kV) between the hollow cathode and anode (deposition chamber body) and are conducted by the capillary towards the target. The impact of the accelerated electron packet (impulse) on the target surface causes the energy transfer from the beam into the target material and, consequently, the target ablation. In other words, the explosion of the target surface in the form of the target material plasma (plume) which propagates in the direction of the substrate where it is deposited (Fig. 1).

The distribution of the electron kinetic energy of the electron beam pulse is rather broad with the majority of electrons gaining the energy around 500 eV.

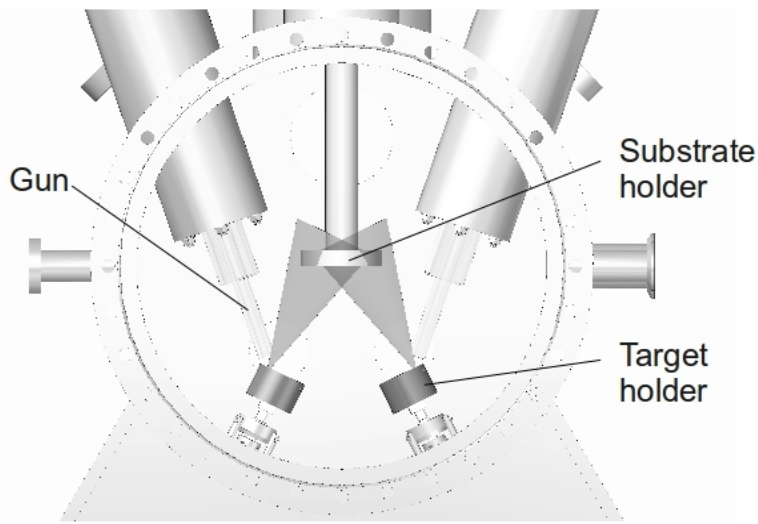

**Fig. 1-Schematic representation of the system for the realisation of the PPD technology.**

The ablation depth and, consequently, the deposition and production rates are determined by the beam energy density, the electron pulse width, the evaporation enthalpy of the target material, its thermal conductivity and by the target density [8-12]. The control of the sample temperature and the in-situ determination of the film thickness are fundamental to achieve successful film growth. These parameters are measured by systems purposely designed by Organic Spintronics, Srl.
The CdTe layers have been deposited using a PPD system equipped by a load-lock chamber. Hence, the external contamination risk is reduced. In addition, the load-lock chamber contains a RF plasma supplier for an in-deep cleaning of the substrates surface (Organic Spintronics, Srl). A linear manipulator makes easy the transfer of the sample from one chamber to the other.
The reactor is equipped with three PPD guns (Organic Spintronics Srl) and three rotating targets holders (Organic Spintronics, Srl). Each target surface is ablated by electrons that come from a specific gun and the angle of inclination of the target surface with respect to the gun axes is 45°.
The CdTe layers have been deposited on quartz and several CdS/ZnO/glass substrates (the CdS and ZnO layers have been deposited via PPD also). The quartz surfaces have been cleaned using acetone and isopropanol baths and finally submitted to an Ar plasma generated by the RF supplier. The CdTe films have been grown under Ar atmosphere with a pressure of $1{,}5\times10^{-3}$ mbar, the gun voltage was kept at 8-12 KV and the repetition rate of electron pulse generation was 10-20 Hz.

The films deposited by PPD have been characterized by XRD, SEM, AFM, Raman and optical spectroscopy methods.
XRD measurements were performed by using a SmartLab-RIGAKU diffractometer equipped with a rotating Cu anode and a parabolic graded-multilayer mirror, placed in front of the sample, to collimate the x-ray beam and to strongly reduce the $K_\beta$ component of the X-ray radiation. The specular diffraction patterns ($\theta/2\theta$ scan) were recorded in the $2\theta$ range of 20÷95°. Raman spectrometer (Horiba-Jobin Yvon T64000) equipped by a krypton laser tuned at 752.5 nm were used [13]. Atomic Force Microscopy measurements (AFM-Smena NT-MDT) in contact mode technique were performed to investigate the morphology of the surface of the CdTe films deposited on quartz substrates. SEM microscope (Leo 1530 ZEISS) equipped with field emission gun as electron source was used for the investigation of the morphology of the surface and cross-section of the deposited layers. Optical spectroscopy (Perkin-Elmer Lamda9) was applied to obtain the optical band gap of the CdTe film deposited on quartz substrate.

# 3. Results and discussion

## 3.1 XRD

Figure 2 shows typical XRD patterns of CdTe/quartz (thin line) and CdTe/CdS/ZnO/quartz (bold line) samples deposited at room temperature. In both systems the CdTe film exhibits a polycrystalline cubic phase (*S.G.= F43m (216), a= 6.48 Å, ICSD #: 043712),* as determined by the analysis of the Bragg peak positions.
The grain sizes of the CdTe, estimated by the FWHM of the peaks applying the Debye-Sherrer equation [14], resulted to be of the same order of magnitude, roughly 30 nm for both samples. The crystallinity of the CdTe film is largely affected by the nature of the underneath layer. The degree of the crystallinity is lower for the CdTe film deposited on the CdS substrate than that one for the CdTe film grown on the quartz substrate, as suggested by the lack of numerous reflections in the pattern of the CdTe/CdS spectra otherwise observed for the CdTe/quartz. The relative peaks intensities are comparable with the ones reported in literature for a random cubic CdTe powder.

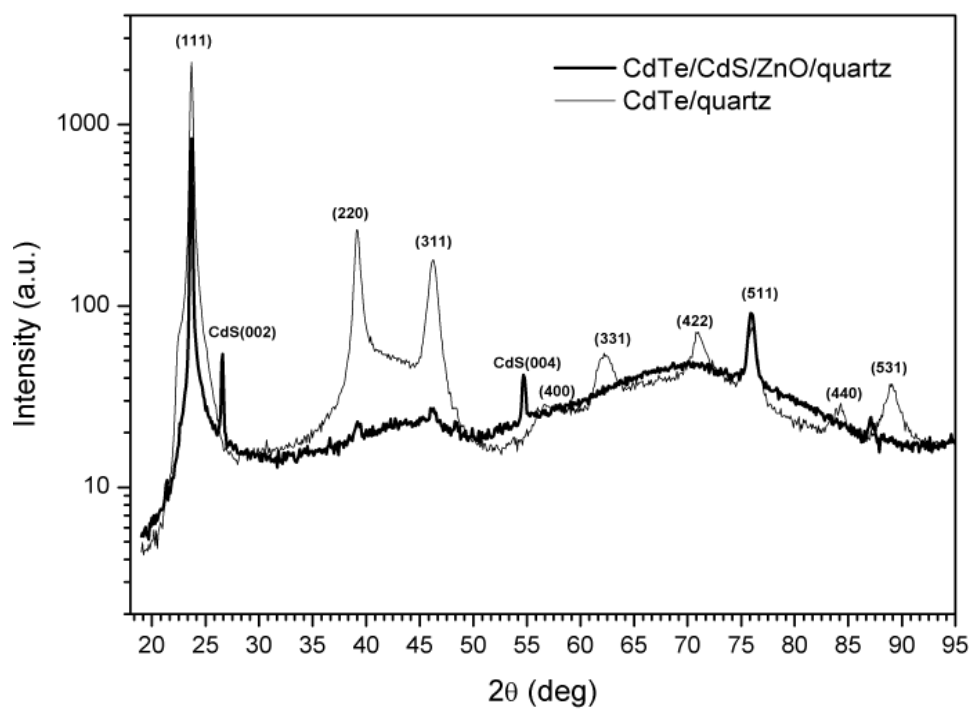

**Fig.2- XRD spectra of CdTe/quartz (thin line) and CdTe/CdS/ZnO/quartz (thick line) samples**

Similar results were obtained [15] using a PLD for the growth of CdTe films; in this case, a polycrystalline structure is observed, but only the (111) and (220) orientations are visible in the spectra. On the other hand, X-ray spectra of electro-deposited CdTe show a (111) preferential orientation, with an increasing crystallinity as a function of the applied potential.
Generally, texturing seems to be present every time films undergo the growth and/or annealing at elevated temperature (e.g. films deposited by close space sublimation technique (CSS) with high substrate temperature [16]). Samples deposited with CSS, PVD (Physical Vapor Deposition) and sputtering techniques show strong orientation in the (111) and (511) directions [17]. This implies that there is also a strong dependence on the underlying substrate structure as well as on the thickness of the films.
On the other hand, the films prepared by the PPD technique exhibit only the formation of a cubic CdTe phase and no preferred orientation of the grains.
The CdS film, in the cell CdTe/CdS/ZnO/quartz, is present as a hexagonal phase (*S.G.= P6₃mc (186), a= 4.14 Å, c= 6.715 ICSD #: 043599, file pdf: 89-2944)*. CdS film is very thin and only few and weak reflections are recorded. A (002) texturing of this layer is clearly present, as documented by the lack of the most intense peak at 28.2° (101) typical for a random powder while the reflections from the planes related to the (002) direction (26.54° and 54.67° respectively) are present. The characteristic dimension of the grains, calculated on the basis of the FWHM of the peaks, is estimated to be around 30 nm.

In the multilayer sample, CdTe is still present as a cubic phase, with a strong texturing along the (111) direction, even though, as mentioned above, the grain sizes of CdTe are smaller than that deposited on quartz. Probably the underneath CdS phase acts as a set of nucleation sites [18].

No peaks related to the ZnO film were detected.

### 3.2 SEM-AFM

The surface of the film is extremely smooth with a measured RMS roughness of 4.6±0.7 nm for the CdTe layer grown on quartz. The low roughness, indispensable for the realization of a multilayer device without structure defects of the layers, indicates besides a homogeneous material deposition also predominantly two-dimensional growth of the layer probably enabled by kinetic energy of the plasma plume. CdTe grains are characterized by using the circularly-averaged autocorrelation function g(r) where the averaged value for the grain's radius is obtained from the position of the first zero-crossing of g(r) (Fig. 3) [19]. The average grain radius measured is 466 ± 23 nm. The absence of periodic peaks in g(r) function suggests a random distribution of the grain dimensions on the surface [20].

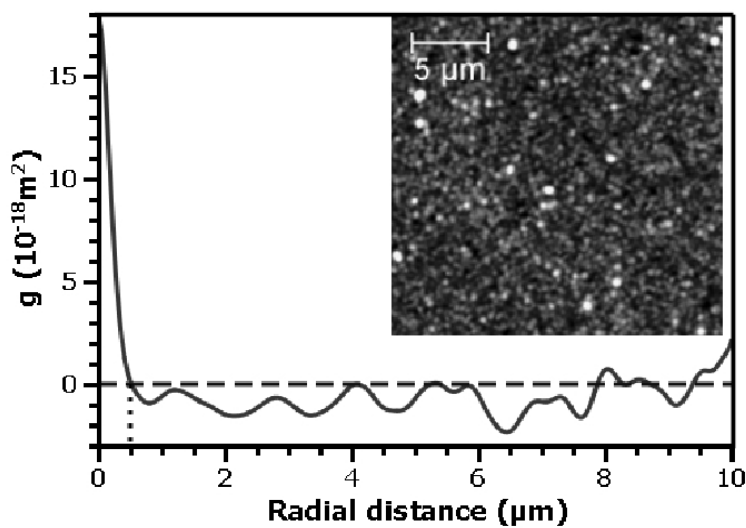

Fig. 3- Circularly-averaged autocorrelation function for the CdTe layer growth on quartz

The discrepancy with the XRD measurements can be understood considering the SEM images of the film surface. As shown in fig.4 the surface is composed of aggregates with different size where the smallest components constituting the aggregates seen also by XRD have average dimension in the range of 30 nanometers. However, the average dimension of the grains present on the surface and seen by both SEM and AFM is of the order of 500 nm. The SEM images show the compact structure of both surface and volume of the film. The surface of the CdTe films grown by sputtering at room temperature shows similar grain size, however the surface of the material is less compact if compared to the surface of CdTe films showed in fig4. CdTe films deposited by CSS shows larger grains due to the necessary high deposition temperature [17].

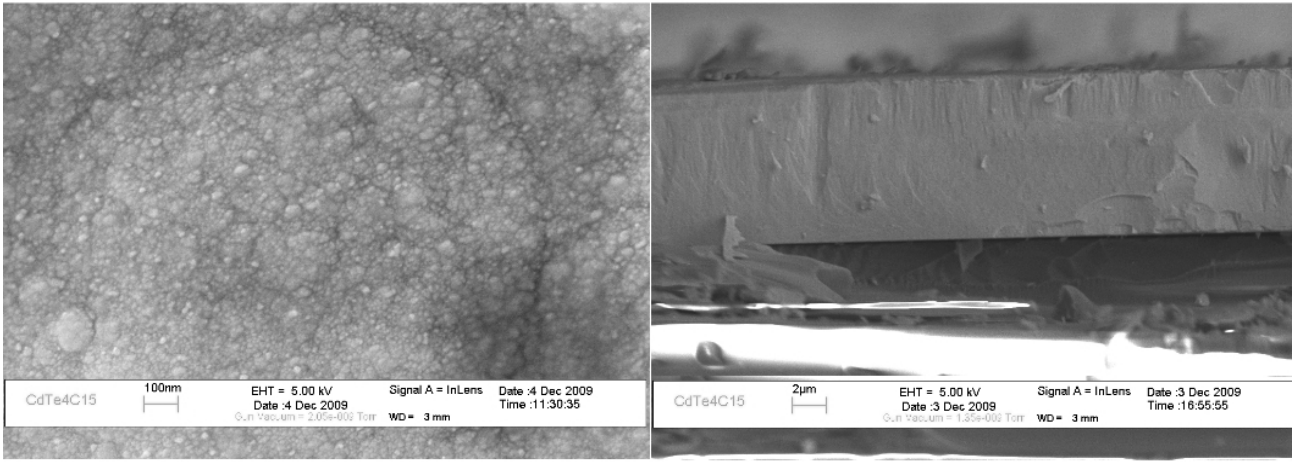

**Fig. 4- Surface (left) and cross-section (right) images of the CdTe film growth on quartz**

The surface of the film deposited on CdS-structure is shown in fig. 5 (an article inherent the deposition of CdS thin films by using the PPD method is in progress). Again, the surface of the film is composed of the aggregates of small particles. However, the dimension of the smallest particles is of the order of 10 nm and the average dimension of the aggregates reaches up to 300 nm. The surface of the film exhibits the same level of compactness as in previous case. The image of the stack of the as deposited CdTe structure is shown in fig. 6. The pieces of the material which appear on the surface of the stack correspond to a debris of the sample fracture.
Each layer of the stack exhibits excellent flatness and a very good interfaces between the layers are observed.

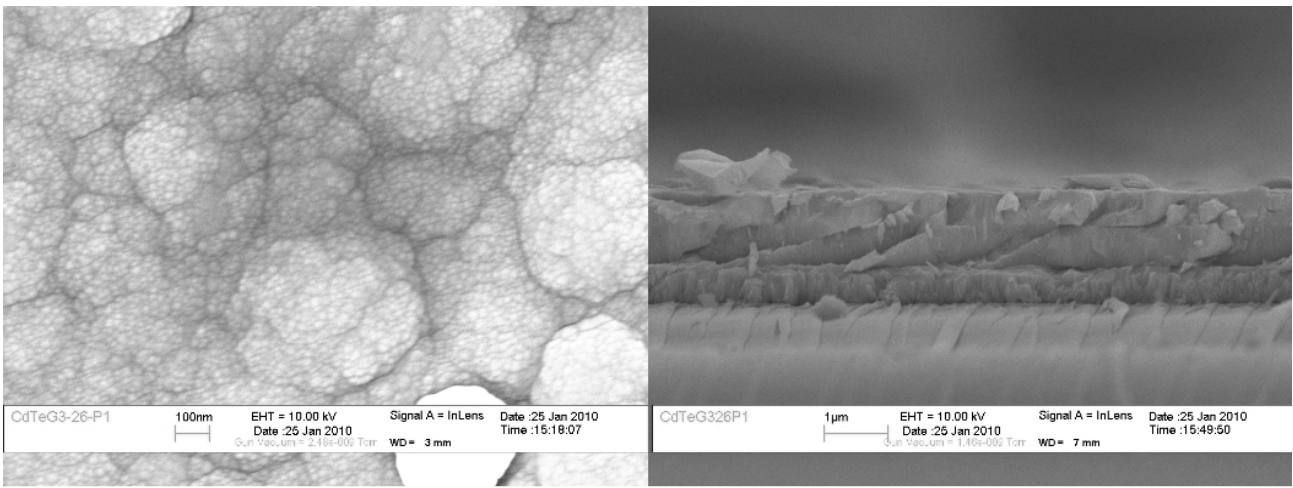

**Fig. 5- Surface (left) and cross-section (right) images of the CdTe film growth on CdS/ZnO/glass**

## 3.3 Optical band edge

The fig. 6 shows the Tauc plot for the CdTe film deposited on quartz. The optical band gap resulting from the optical analysis of the film is 1.50 eV. This value is in excellent agreement with the one reported in the literature for the bulk cubic phase of CdTe at 300 K [21] indicating that, although the surface and the bulk of the film is constituted of nano-grains (as reported from the XRD and SEM study), the grain borders apparently contain rather small amount of structural
 defects.

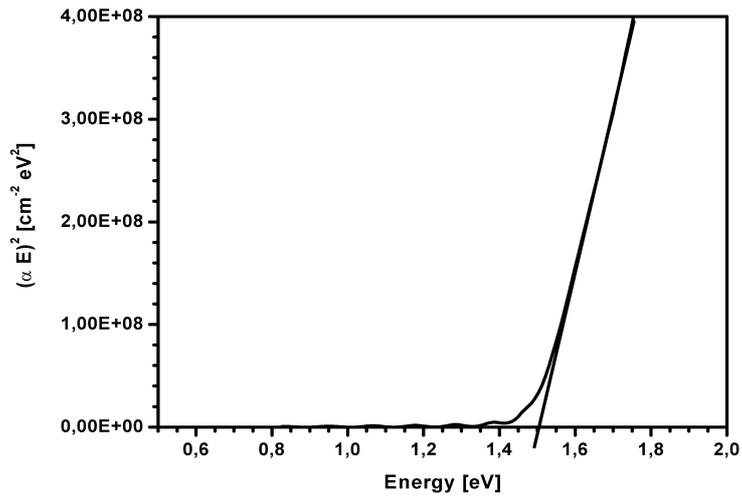

**Fig.6- Tauc plot of the CdTe film deposited on quartz**

### 3.4 Raman spectra

The Raman spectrum (fig. 7), obtained using the 752.5 nm excitation wavelength, shows 3 peaks at 122, 140 and 165 cm$^{-1}$.

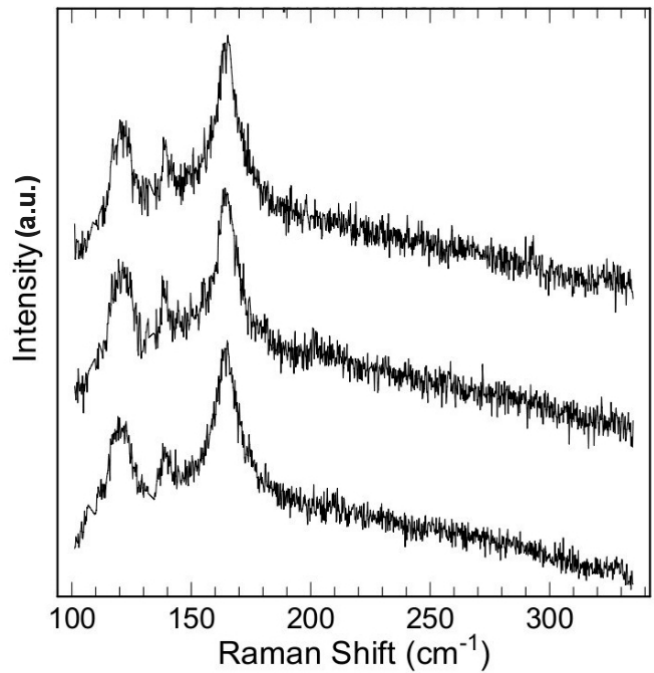

**Fig. 7- Raman spectra ($\lambda_{exc}$ = 752.5 nm) for three different points on the surface of a CdTe/quartz sample**

The highest energy peaks at 165 cm$^{-1}$ and at 140 cm$^{-1}$ correspond to the longitudinal (LO) and to the transverse (TO) modes, respectively, of the CdTe bulk crystal [22, 23], whereas the peak at 122 cm$^{-1}$, is related to the TeO$_2$ Raman vibrational mode [22]. The entire film area was tested point by point, showing no significant changes among the different spectra, thus indicating a good homogeneity of the deposited film.

# 4. Conclusions

The properties of the as-deposited films have been studied and the results reveal their exceptional properties as e.g. the cubic phase, the morphology dominated by very small particles ($\leq$ 30 nm) and extremely flat and compact surface (RMS roughness of 4.6±0.7 nm). The value of 1.50 eV for the optical band gap is in excellent agreement with the literature data. Such results confirm that the PPD technique is excellent tool for large-area low-cost thin film solar cells preparation at industrial scale. Siena Solar Nanotech is presently developing, with the close collaboration of Organic Spintronics, the sources and tools to allow the transfer of these results from the laboratory based deposition towards the wide area web based industrial prototype that may integrate into the thin film fabrication line.

**Acknowledgement**

Authors acknowledge the "Flexsolar Project", part of the "Industria 2015" plan for the economic development by the italian "Ministero dello Sviluppo Economico" for the strong financial support to this work.